\documentclass[referee]{2_aa}
\usepackage{graphics, epsfig}
\begin{document}

\title{Cosmological Black Holes as Seeds of Voids in Galaxy Distribution}

\author{S. Capozziello\inst{1,4}
        \and M. Funaro\inst{2}
        \and C. Stornaiolo\inst{3,4}}

\offprints{capozziello@sa.infn.it}

\institute{Dipartimento di Fisica ``E.R. Caianiello'',
Universit\`a di Salerno, Via S. Allende, 84081 - Baronissi
(Salerno), Italy \and Dipartimento di Matematica e Informatica,
Universit\`{a} di Salerno, Via Ponte Don Melillo, 84084 - Fisciano
(Salerno), Italy \and Dipartimento di Scienze Fisiche,
Universit\`{a} di Napoli, Complesso Universitario di Monte S.
Angelo, Via Cinthia, Edificio N - 80126 Napoli, Italy \and
Istituto Nazionale di Fisica Nucleare, Sezione di Napoli,
Complesso Universitario di Monte S. Angelo, Via Cinthia, Edificio
G -  80126 Napoli, Italy}

\date{Receveid / Accepted }

\titlerunning{CBH as seeds of voids}

\maketitle

\begin{abstract}

Deep surveys indicate a bubbly structure of cosmological large
scale which should be the result of evolution of primordial
density perturbations. Several models have been proposed to
explain origin and dynamics of such features but, till now, no
exhaustive and fully consistent theory has been found. We discuss
a model where cosmological black holes, deriving from primordial
perturbations, are the seeds for large-scale-structure voids. We
give details of dynamics and accretion of the system
voids-cosmological black holes from the epochs $(z\simeq10^{3})$
till now finding that void of $40h^{-1}Mpc$ of diameter and
under-density of $-0.9$ will fits the observations without
conflicting with the homogeneity and isotropy of cosmic microwave
background radiation.


\end{abstract}

\section{\normalsize Introduction}
The existence of voids has been  evident after the discovery by
Kirshner et al. of a large void with diameter of $60$ Mpc in
B\"{o}otes (\cite{kirsh}). Systematic surveys have shown the
existence of many regions with similar characteristics. Computer
analysis of galaxy distribution gives evidence that voids occupy
about 50\% of the volume of the universe (e.g. see
\cite{El-Ad:1997af}) or, according to a more recent paper
(\cite{hoyle}), about $40$\% of the volume of the universe.

Today, there is a large agreement on the issue that voids are not
just empty regions of space, but that they are regions with a very
low density of luminous matter.

As observed by Peebles (\cite{peebvoid}), the low dispersion of
velocities of galaxies indicates that most of matter must be
inside the voids, not only if the density parameter (for the
matter component) is $\Omega_{m}=1$ but also in the case
$\Omega_{m}\ll 1$. In any case, recent observations
(\cite{boomerang}; \cite{supernovae1}; \cite{supernovae2})
indicate that the total value of density parameter is $\Omega=
\Omega_{m}+ \Omega_{\Lambda}=1$ where $\Omega_{m}\simeq 0.3$ and
$\Omega_{\Lambda}\simeq 0.7$. In this case $\Omega_{\Lambda}$ is
the contribution due to the whole content of unclustered matter
which can be cosmological constant, some kind of scalar field
(\cite{quintessence1}; \cite{quintessence2}; \cite{quintessence3};
\cite{quintessence4}) or, in general, "dark energy".

It is worthwhile to stress that the visual inspection of galaxy
distribution suggests nothing else but the absence of large amount
of luminous matter in wide regions. Furthermore, it is not clear
whether the voids are spherical regions approximately empty or
under-dense regions with arbitrary shapes. Several definitions of
voids have been proposed, but a general agreement on their real
nature has not been reached yet (\cite{sch}).

The Swiss-Cheese cosmological model,  initially proposed  by
Einstein and Straus (\cite{einstraus1}; \cite{einstraus2}),
appears suitable for the description of the cosmological voids. In
a recent paper (\cite{cosimo}), it was proposed an approach for
the formation of the cosmological voids in the framework of such
model. It was shown that voids are the consequence of the collapse
of extremely large wavelength perturbations into low-density black
holes and of the comoving expansion of matter surrounding the
collapsed perturbations.

As a result, it was claimed that in the center of each void there
is a black hole whose mass $M$ compensates the mass which the void
would have if it were completely filled with matter having a
cosmological density.

In \cite{cosimo}, the voids are empty regions of the universe
which grow comovingly with the cosmological expansion. In that
case, the presence of cosmic background radiation was neglected.

In this paper, we analyze the physical mechanism capable of
explaining the structure of voids in presence of baryonic matter,
cosmic background radiation (CBR) with central black holes acting
as seeds. The layout of the paper is the following: in Sect.2, we
will present the cosmological black hole (CBH) model in the
framework of the Friedmann-Lema\^{i}tre-Robertson-Walker (FLRW)
cosmology. Sect.3 is devoted to the discussion of the effects of
interaction between the CBR and CBH. A mechanism for the formation
of an under-density regime void is analyzed in Sect.4, while the
matching with observations, which allows to determine the initial
time of voids formation and the mass function of CBHs, is studied
in Sect.5. The discussion of results and conclusions are given in
Sect.6.

\section{\normalsize The CBH model and Cosmology}

The cosmological model proposed in \cite{cosimo} is an
Einstein-Straus universe which is embedded in a FLRW metric. A
central spherical black hole with mass
\begin{equation}\label{massflat}
 M=\frac{4}{3}\pi \Omega_{CBH}\rho_{c} R_{v}^{3},
\end{equation}
is present in all the voids.

In Eq. (\ref{massflat}), the parameter
\begin{equation}\label{omegacbh}
\Omega_{CBH}=\frac{\rho_{CBH}}{\rho_{c}}
\end{equation}
represents the fraction of density due to all these black holes
with respect to the total density of the universe; $\rho_{c}
=1.88\times 10^{-29}\,g\, cm^{-3}h^{2}\Omega$ is the today
critical density of the universe.

All the voids are assumed to be spherical.

A black hole forms when a body of mass $M$ collapses entirely
within a sphere of radius

\begin{equation}\label{schw}
R_s=\frac{2GM}{c^2}.
\end{equation}
This statement is equivalent to say that its density satisfies the
relation,
\begin{equation}\label{schwdens1}
R_s(\rho)= \sqrt{\frac{3c^2}{8\pi G\rho}}.
\end{equation}
Conversely,  Eq.(\ref{schwdens1}) relates to any density  a
corresponding Schwarzschild radius. In other words, any space-like
sphere of matter, with uniform density $\rho$ and radius equal to
$R_s(\rho)$, is a black hole.

From Eqs.(\ref{massflat}) and (\ref{schw}), we determine the mass
$M$ and the corresponding Schwarzschild radius and consequently,
from Eq.(\ref{schwdens1}) the mass density of the central black
hole. For example, a black hole in the center of a $50h^{-1}$ Mpc
diameter void would have a mass $1.8\times10^{16} M_{\odot}$
corresponding to a Schwarzschild radius $1.7$ kpc and a density of
$2.34\times10^{-16}gcm^{-3}$.

The above value of density is the one reached by the collapsing
matter when it crossed the Schwarzschild radius.  It suggests that
the process of formation of a CBH started with large wavelength
perturbations at cosmological densities of the order of
$10^{-19}g/cm^{3}$ i.e for $1+z \approx 10^{3}$. According to the
inflationary scenario, we only need to suppose that the inflation
occurred during a time long enough to provide such perturbations
\footnote {The Oppenheimer-Snyder model (\cite{opsny}) describes
the spherical symmetric collapse at zero pressure.}.

It is important to note that, since the Einstein-Straus model
assumes spherical symmetry, the perturbation  does not experience
the cosmic expansion during its collapse.

For this reason, we can assume that the total mass of perturbation
\begin{equation}\label{totalmass}
M_{p}=\frac{\pi}{6} \Omega_{p}\rho_{i} \lambda_i^{3}
\end{equation}
remains constant during the whole process.

The Schwarzschild radius of the spherical perturbation is equal to

\begin{equation}\label{raggio}
 R_{s}=\frac{H_i^2}{c^2}\left(\frac{\lambda_i}{2}\right)^3\Omega_p.
\end{equation}

We can distinguish two cases. The case in which the relation
\begin{equation}\label{raggiomax}
\frac{2 R_{s}}{\lambda_{i}}\geq1,
 \end{equation}
is satisfied.  The perturbation is in linear regime  and,
according to the evolution equations of a universe with perfect
fluid equation of state, it is frozen when $\lambda_{i}$ is larger
than the Hubble radius (\cite{mukhanov}). After crossing the
Hubble horizon, it collapses and becomes a black hole when
\begin{equation}\label{limiti}
\left(\frac{ \lambda_{i}}{2}\right)^{2}\geq
\frac{c^{2}}{H_{i}^{2}\Omega_p}.
\end{equation}
In the case
\begin{equation}\label{raggiomin}
\frac{2R_{s}}{\lambda_i} <1,
\end{equation}
the perturbation evolves as shown in \cite{cosimo}. During the
contraction, it becomes unavoidably a black hole if the final
density is very low and the internal pressure and temperature
cannot raise to values large enough to prevent the collapse. This
is true also when the perturbation enters in a non-linear regime.
In addition, any possible centrifugal barrier can be reduced to
values smaller than the Schwarzschild radius (\cite{loeb}) by the
interaction of collapsing matter with the cosmic background
radiation. According to the Einstein-Straus model, after the
formation of black hole, the matter around it expands in a
comoving way leading to the formation of an empty region between
it and the rest of the universe.  As the central black hole cannot
be seen, the whole region appears as a void to an external
observer.

A CBH can be detected through its lensing properties, since it
behaves like a Schwarzschild  gravitational lens. According to our
hypothesis, a CBH sits in the center of a void and the Einstein
angle is (\cite{sef})
\begin{equation}\label{angle}
\alpha_{0}=4.727\times
10^{-4}\Omega_{CBH}^{1/2}\sqrt{\frac{R_{V}^{3}D_{ds}}{D_{d}D_{s}}},
\end{equation}
where $R_{V}$ is the radius of the void, $D_{s}$ is the distance
of the source from the observer, $D_{ds}$ is the distance of the
source from the CBH, and $D_{d}$ is the distance of the CBH from
the observer, all these quantities are expressed in Mpc. For a 50
Mpc void of diameter, with the center placed at a distance of 80
Mpc from the Sun and with the source at the opposite edge of the
void, we expect a deflection angle $$\alpha_{0}\simeq 3.2\times
10^{-4}\Omega_{CBH}^{1/2}. $$ On the other hand, Zeldovich and
Sazhin point out in (\cite{zelsa}), that generally static
structures can raise the temperature of the cosmic background
radiation by an amount proportional to the Hubble parameter and
the gravitational time delay. By considering the Swiss-Cheese
model case, they find a fluctuation of temperature $\delta T/T
\sim 10^{-10}$ for a giant galaxy ($M= 4\times10^{12} M_{\odot}$).
Since this result is proportional to the mass, it follows that it
corresponds to a fluctuation of temperature $\delta T/T \sim
10^{-5}$ for a large CBH with mass ($M\sim 10^{17} M_{\odot})$,
which does not contradict the recent Boomerang and WMAP
measurements (\cite{boomerang}; \cite{wmap}).

An interesting feauture of voids with central CBH is derived
starting from a Swiss-Cheese model based on an Einstein-de Sitter
cosmology. To this aim, let us consider the energy balance in
Newtonian terms for a galaxy with mass $m$ sitting on the edge of
a void i.e at a distance $r=r_{e}a(t)$ where  $r_{e}$ is a
covariant radius. We have
\begin{equation}\label{energy}
E=\frac{1}{2}m v^2 - G\frac{mM_{v}}{r_{e}a}.
\end{equation}

As the void is comovingly expanding, we can impose $v=Hr_{e}a(t)$.
The mass $M_{v}$ inside the void is
\begin{equation}\label{2}
    M_{v}=\frac{4\pi\rho_{int} (r_{e}a)^{3}}{3}.
\end{equation}

By Eq. (\ref{energy}), one obtains
\begin{equation}\label{3}
E=\frac{m(r_{e}a)^{2}}{2}\left[H^{2} -\frac{8\pi
G\rho_{int}}{3}\right]
\end{equation}
and then
\begin{equation}\label{3b}
E=\frac{m(r_{e}a)^{2}}{2}\left[ \frac{8\pi
G}{3}(\rho_{tot}-\rho_{int})\right].
\end{equation}
From this relation, we see that, $E=0$ if
 $\rho_{tot}=\rho_{int}$, i.e. the expansion velocity of a galaxy on the edge of a void coincides
with the escape velocity\footnote {In the densities $\rho_{int}$
and $\rho_{tot}$, we are taking into account all the contributions
to the energy density as matter and radiation.}. On the other
hand, a galaxy on the edge of a void is gravitationally bounded
with a CBH, if $\rho_{int}>\rho_{tot}$. In this paper, we shall
adopt as \textit{definition} of edge of a void the spherical
region where the galaxies have energy $E=0$.

These considerations can be immediately extended to the case in
which the spatial curvature is different from zero (i.e. to
Friedmann models different from the considered Einstein-de Sitter
one). This simply implies that
\begin{equation}\
H^{2}=\frac{8\pi G}{3}\rho_{tot}-\frac{K}{a^{2}}
\end{equation}
and Eq.(\ref{3}) becames
\begin{equation}\
E=\frac{m(r_{e}a)^{2}}{2}\left[ \frac{8\pi
G}{3}(\rho_{tot}-\rho_{int})-\frac{K}{a^{2}}\right].
\end{equation}
In this case, the value of $E$ depends also on $K/a^{2}$.

\section{\normalsize Interaction between CBR and CBH}
\label{interaction}

So far we have considered the Einstein-Straus Swiss-cheese model
in a universe filled only with matter. In this section, we shall
show how the interaction with CBR will lead to an accretion of
CBH. Since the pressure of radiation is different from zero, the
radiation itself may cross the edge of the voids regardless the
Einstein-Straus junction conditions (\cite{einstraus1};
\cite{einstraus2}). As we consider this physical process during
the matter epoch, we can neglect the contribution of radiation to
the black hole formation.

A CBH absorbs energy from CBR according to the law
(\cite{custodio})

\begin{equation}\label{accretion}
    \frac{d\, M}{d\,t}=\sigma_{g}(M) F_{rad}
\end{equation}
where $\sigma_{g}(M)=(27\pi G^{2}/c^{4})M^{2}$ is the
gravitational cross section of CBH and $F_{rad}=\rho_{rad}c$ is
the radiation flux of CBR.

In a matter dominated universe, (dust), the evolution of radiation
density is given  by
\begin{equation}\label{evraddens}
\dot{\rho}_{rad}= -3H(\rho_{rad}+p_{rad})
\end{equation}
where ${\displaystyle H=\frac{2}{3t}}$ and the equation of state
is ${\displaystyle p_{rad}=\frac{1}{3}\rho_{rad}}$.

Immediately, we get
\begin{equation}\label{rho}
\rho_{rad}(t)= \frac{A}{t^{\frac{8}{3}}},
\end{equation}
where the constant $A$ has the dimensions
$[gr\,cm^{-3}sec^{8/3}]$. At present epoch, assuming $t_{0}\cong
3.08\times10^{17}h^{-1} sec$ (this value is consistent with WMAP
observations (\cite{wmap}) for $h\cong 0.7$) and
$\rho_{rad}(t_{0})\cong 4.8 \times 10^{-34}gr\,cm^{-3}$,  we
obtain

\begin{equation}\label{scaleofrho}
\rho_{rad}(t) \cong \frac{2.08\times
10^{13}h^{-8/3}}{t^{\frac{8}{3}}},
\end{equation}
dropping the physical dimensions.

Eq.(\ref{accretion}) becomes
\begin{equation}\label{accretion2}
\frac{d M}{d\,t}=\frac{2.94\times
10^{-31}h^{-8/3}M^{2}}{t^{\frac{8}{3}}}.
\end{equation}

Integrating from an initial epoch $t_{i}$ to the present epoch, we
have that the ratio between the present mass $M_0$ and the initial
mass $M_i$ is

\begin{equation}\label{mass}
   M_{0}   = \frac{M_{i}}{1+  1.76\times 10^{-31}h^{-8/3}M_{i}\left(
    \frac{1}{t_{0}^{5/3}}-\frac{1}{t_{i}^{5/3}}\right)}.
\end{equation}

If $t_{i}\ll t_{0}$, we can approximate the formula for the mass
accretion to

\begin{equation}\label{approx}
 M_{0} = \frac{ M_{i}}{1-
    \frac{1.76\times 10^{-31}h^{-8/3}M_{i}}{t_{i}^{5/3}}}.
\end{equation}

The growth of mass of a CBH is then
\begin{equation}\label{increment}
\Delta M=M_{0}-M_{i}=M_{i}\left[\frac{1.76\times
10^{-31}h^{-8/3}M_{i}\left(\frac{1}{t_{0}^{5/3}}-\frac{1}{t_{i}^{5/3}}\right)}{1+
1.76\times
10^{-31}h^{-8/3}M_{i}\left(\frac{1}{t_{0}^{5/3}}-\frac{1}{t_{i}^{5/3}}\right)}\right].
\end{equation}

For $t_{i}\ll t_{0}$, we have
\begin{equation}\label{moreapprox}
\frac{M_{0}}{M_{i}}\simeq
 1+\frac{1.76\times 10^{-31}h^{-8/3}M_{i}}{t_{i}^{5/3}}.
\end{equation}

Therefore, it is evident that the BH accretion is proportional to
its initial mass.

\section{\normalsize Under-density inside the voids}

From the point of view of density, voids are under-dense regions
of space depleted of galaxies with respect to the external
background.  We can estimate the amount of such an under-density
in the framework of our CBH-void model.

Taking into account the accretion process in the void, we cannot
neglect the fact that, as the void increases in volume, a certain
amount of galaxies enter in the void contributing to the total
mass inside the void. This implies that the accretion of mass
inside the void, given by (\ref{accretion}), has to be corrected
as

\begin{equation}\label{accretionv}
    \frac{d\, M}{d\,t}=2\sigma_{g}(M) F_{rad}.
\end{equation}

Factor $2$ can be  justified in the following way.

If we consider any mass increment in Eq.(\ref{energy}), it is easy
to observe that the mass of galaxies entering in the void is equal
to the mass increment of the black hole, if one takes into account
the conservation equation for $\rho_{tot}$. In first
approximation, one can double the righthand side of
Eq.(\ref{accretion}). To be rigorous, one should consider a delay
effect, since gravity does not propagates instantaneously. Due to
this fact, it would be correct to deal with this problem under the
standard of the General Relativity.  However, it is possible to
show that the delay effect can be neglected, at least in first
approximation.

To find the density of galaxies $\rho_{und} $ which enter to
expanding voids, one has to subtract the accreted mass of CBH,
obtained from Eq.(\ref{approx}), to the total final accretion
mass, given by Eq.(\ref{accretionv}), (in the approximation
$t_{i}\ll t_{0}$)

\begin{equation}\label{approxvoid}
   M_{f}   = \frac{M_{i}}{1-  3.53\times 10^{-31}h^{-8/3}M_{i}t_{i}^{-5/3}},
\end{equation}
 and divide the result by the volume of the void taken as
$M_{f}/\rho_{tot}$.

Finally, we find that

\begin{equation}\label{rhound}
    \rho_{und}=\left(1-\frac{M_{0}}{M_{f}}\right)\rho_{tot}
\end{equation}

The under-density is given by the contrast of density of galaxies
in the void with the total density $\rho_{tot}$, i.e.
\begin{equation}\label{underdensity}
   \delta=\frac{\rho_{und}}{\rho_{tot}}-1=-\left(1+\frac{9.69\times
   10^{16}h^{-2/3}V}{t_{i}^{5/3}}\right)^{-1}\,,
\end{equation}
where $V$ is the volume of the void expressed in $10^{3}Mpc^{3}$
and $t_{i}$ is the initial time expressed in seconds. From this
last equation, we obtain the formula for the initial time $t_{i}$,
which is
\begin{equation}\label{ti}
   t_{i}=\left(  -\frac{\delta}{1+\delta}\times 9.69\times
   10^{16}h^{-2/3}V\right)^{\frac{3}{5}},
\end{equation}
where $\delta$ and $V$ can be retrieved from the respective values
found in the catalogs (\cite{El-Ad:1997af}). On the other side we
can also determine the initial masses of the CBHs just after their
formation. This can be expressed by the formula
\begin{equation}\label{mi}
M_{i}=\frac{\rho_{int} V}{1-2\frac{1+\delta}{\delta}}
\end{equation}
where $\rho_{int}$ is the internal density of the void.

\section{\normalsize Matching with the observations}
Using the data of the volume and the under-density of  voids given
in the catalog of El-Ad and Piran (\cite{El-Ad:1997af}) which
deduced their data from SSRS2 (\cite{SSRS2_1}; \cite{SSRS2_2}) and
IRAS (\cite{IRAS}) observations, we found the following initial
times and corresponding initial and final masses according to
formulas (\ref{ti}, \ref{mi}, \ref{approxvoid}). \vskip1truecm

{\center{
\begin{tabular}{|c|c|c|c|c|}
      \hline
     Volume  & under-density & initial time & initial CBH masses &final CBH  masses\\
        $10^{3}Mpc^{3}$& $\delta$& $ h^{-2/5} 10^{12}$ sec &  $ \Omega h^210^{16}M_\odot$ &  $\Omega h^210^{16}M_\odot$ \\
      \hline

      84.9 & -0.89 &0.78 & 1.90&  2.11\\
      93.9 & -0.87 & 0.74 & 2.01&  2.28 \\
      119.0& -0.93 & 1.29 & 2.88 &  3.08\\
      24.0 &-0.91 & 0.42  & 0.56&  0.61\\
      22.6 & -0.94 & 0.53 & 0.56&  0.59 \\
      17.4 & -0.92 &  0.37 & 0.41&   0.47\\
      8.8 & -0.91 &  0.23 & 0.20& 0.22\\
      11.4 &  -0.95 & 0.39 & 0.29& 0.30\\
      31.1 & -0.86 & 0.36  & 0.65& 0.74\\
      22.4 &  -0.69 & 0.16  & 0.33& 0.43\\
      41.5 & -0.88 & 0.48  & 0.91& 1.02\\
       8.8 & -0.97 & 0.46  & 0.23& 0.24\\
      10.7 &  -0.74 & 0.12  & 0.17& 0.22\\
      \hline
    \end{tabular}
    }
}

\vskip1truecm  {\bf Table 1}: {\small Evolution of voids and black
holes features derived from our model using the data in
\cite{El-Ad:1997af}.}

\vspace{2. mm}

Dynamics deduced from our model is consistent with observations
and seems to confirm the gravitational origin of  voids. In this
approach, the role of dark matter has to be revised since most of
the mass (about one half) of the structure is concentred in the
central CBH. In this picture, it is only the density contrast
between voids and background which drives dynamics. CBHs are just
the remnant of primordial collapsed perturbations while voids, or
precisely the edges of voids, are the result of perturbations
which wavelength following the cosmic expansion. The whole system,
also if expanding and interacting with CBR, remains in
equilibrium.

\section{\normalsize Discussion and Conclusions}
In this paper, we have developed a model where cosmological black
holes are seeds for large scale structure voids. Such systems come
out from the evolution of primordial perturbations and result as
stable structures from $(z\simeq10^{3})$ up to now. They enlarge
till diameters of about $40h^{-1}Mpc$ and the under-density of
voids is of the order $-0.9$ with respect to the background. The
whole structure is a sort of honeycombs where most of galaxies
(i.e. luminous matter) are located on the edge of voids while most
of dynamical mass is sited in the central black hole. The edge is
defined by a natural equilibrium condition on the energy due to
the balance of gravitational pull of the central black hole and
the cosmic expansion. The cosmic background radiation contributes
to the accretion of black hole flowing inside the void but its
homogeneity and isotropy is not affected in agreement with data.
The picture which emerge agrees with optical and IRAS observation
(\cite{El-Ad:1997af}) giving a $\sim50\%$ of the volume filled by
voids with the above characteristics. The presence of central
black holes seems to confirm the gravitational origin of the voids
and stabilizes the system against cosmic expansion preventing its
evaporation.

It is interesting to note that the order of magnitude observed for
the masses of CBH concides with the one of the Great Attractor
(\cite{Fairall}). It is very tempting for us to identify the Great
Attractor as a CBH and to use the model described in this paper
for explaining the large scale motions observed for the galaxies
surrounding it.

However, the main problem with observations of cosmic velocity
fields (\cite{Faber}) is that the voids are, in general, the
contrary to Great Attractor, and large scale structure around the
voids do not show velocity fields converging toward the voids, but
toward the visible clusters and superclusters around the voids.
This apparent shortcoming, in the framework of our model (see also
\cite{cosimo}), can be overcome by the Birkoff theorem which
states that the stationary solutions are also static if the
spherical symmetry is restored. So, a fraction of galaxies is
attracted by clusters and superclusters "outside" the void while
another fraction has no dynamics since it has been already
attracted "inside" the void. This fact could be interpreted as an
early selection due to a competitive mechanism between CBHs and
external matter contained into clusters and superclusters.

However, if the Swiss-Cheese model were always valid such a
selection would have never been achieved; instead, in a more
realistic situation, the model holds only approximately so then we
have to expect galaxies inside and outside the void due to the
deviations from sphericity and to the perturbations of the CBH
mass.

Furthermore, as observed in \cite{Davis} and \cite{peebvoid}, the
small relative velocity dispersion shows that, if $\Omega_{M}=1$,
then most of the mass has to be contained into the voids. The same
authors conjecture that this must be true even when $\Omega_{M}$
is smaller than 1 as predicted by several CDM simulations.

\end{document}